\documentclass[12pt]{article}
\usepackage{graphicx}

\newcommand{\Lya}{Ly$\alpha$~}

\def\napoli{
  Raymond and Beverly Sackler School of Physics and Astronomy\\
  Tel Aviv University, Tel Aviv 69978, Israel}
\def\support{\footnote{This work was supported by Israel Science
    Foundation grant 823/09 and the Ministry of Science and
    Technology, Israel.}}

\def\Title#1{\begin{center} {\Large #1 } \end{center}}
\def\Author#1{\begin{center}{ \sc #1} \end{center}}
\def\Address#1{\begin{center}{ \it #1} \end{center}}

\newenvironment{Abstract}{\begin{quotation}  }{\end{quotation}}
\newenvironment{Presented}{\begin{quotation} \begin{center} 
             PRESENTED AT\end{center}\bigskip 
      \begin{center}\begin{large}}{\end{large}\end{center} \end{quotation}}
\def\Acknowledgements{\bigskip  \bigskip \begin{center} \begin{large}
             \bf ACKNOWLEDGEMENTS \end{large}\end{center}}

\def\beq{\begin{equation}}
\def\eeq#1{\label{#1}\end{equation}}
\def\eeqn{\end{equation}}

\def\beqa{\begin{eqnarray}}
\def\eeqa#1{\label{#1}\end{eqnarray}}
\def\eeqan{\end{eqnarray}}

\let\bar=\overbar

\def\Dslash{\not{\hbox{\kern-4pt $D$}}}
\def\dslash{\not{\hbox{\kern-2pt $\del$}}}

\def\msb{{\bar{\ssstyle M \kern -1pt S}}}

\begin{document}
\begin{titlepage}

\vfill
\Title{Detecting the First Stars and Black Holes with 21-cm Cosmology}
\vfill
\Author{ Rennan Barkana\support}
\Address{\napoli}
\vfill
\begin{Abstract}
  Understanding the formation and evolution of the first stars and
  galaxies represents one of the most exciting frontiers in astronomy.
  Since the universe was filled with neutral hydrogen at early times,
  the most promising method for observing the epoch of the first stars
  is using the prominent 21-cm spectral line of the hydrogen atom.
  Current observational efforts are focused on the reionization era
  (cosmic age $t\sim 500$ Myr), with earlier times considered much
  more challenging. However, the next frontier of even earlier galaxy
  formation ($t\sim 200$ Myr) is emerging as a promising observational
  target. This is made possible by a recently noticed effect of a
  significant relative velocity between the baryons and dark matter at
  early times. The velocity difference significantly suppresses star
  formation. The spatial variation of this suppression enhances
  large-scale clustering and produces a prominent cosmic web on 100
  comoving Mpc scales in the 21-cm intensity distribution. This
  structure makes it much more feasible for radio astronomers to
  detect these early stars, and should drive a new focus on this era,
  which is rich with little-explored astrophysics.
\end{Abstract}
\vfill
\begin{Presented}
The 10th International Symposium 
on Cosmology and Particle Astrophysics (CosPA2013)\\
Honolulu, Hawaii, U.S.A., November 12--15, 2013
\end{Presented}
\vfill
\end{titlepage}
\def\thefootnote{\fnsymbol{footnote}}
\setcounter{footnote}{0}

\section{Introduction}

Galaxies around us have been mapped systematically out to a redshift
$z \sim 0.3$ by recent large surveys. The observed galaxy distribution
shows a large-scale filament-dominated ``cosmic web'' pattern that is
reproduced by cosmological numerical simulations. This structure is
well-understood theoretically as arising from the distribution of the
primordial density fluctuations, which drove hierarchical structure
formation in the early universe. Recent observations have been pushing
a new frontier of early cosmic epochs, with individual bright galaxies
detected reliably from as early as $z = 7.2$ \cite{z7p2}, which
corresponds to $t \sim 750$ Myr after the Big Bang.

\begin{figure}[htb]
\centering
\includegraphics[width=142mm]{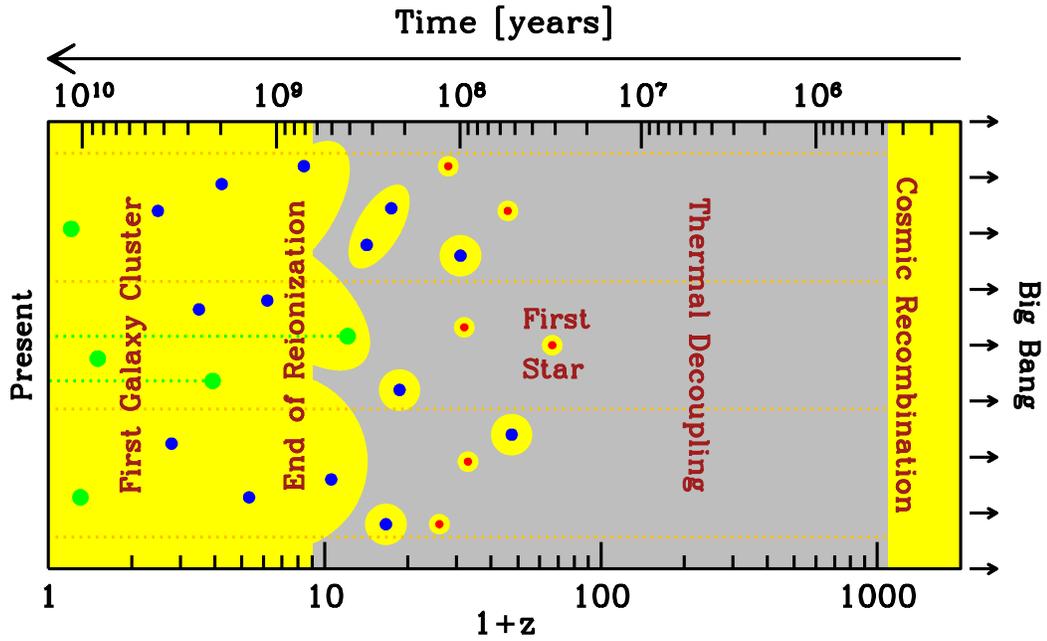}
\caption{Overview of cosmic history, with the age of the
universe shown on the top axis and the corresponding redshift on the
bottom axis. Yellow represents ionized hydrogen, and gray is
neutral. Observers probe the cosmic gas using the absorption of
background light (dotted lines) by atomic hydrogen. Stars formed in
halos whose typical size continually grew with time, going from the
first generation that formed through molecular-hydrogen cooling (red
dots), to the larger galaxies that formed through atomic cooling and
likely dominated cosmic reionization (blue dots), all the way to
galaxies as massive as the Milky Way, some of which host bright
quasars (green dots). From \cite{BSc}.}
\label{f:history}
\end{figure}

The best hope of observing the bulk population of early stars is via
the cosmic radiation fields that they produced. In particular, this
radiation affected the hyperfine spin-flip transition of neutral
hydrogen (H~I) at a wavelength of 21 cm, potentially the most
promising probe of the gas and stars at early times. Observations of
this line at a wavelength of $21\times (1+z)$ cm can be used to slice
the universe as a function of redshift $z$ and obtain a
three-dimensional map of the diffuse H~I distribution within it
\cite{Hogan}, in the previously unexplored era of redshifts $\sim 7 -
200$. See Figure~\ref{f:history} for a brief summary of early cosmic
history.

A major goal of current and upcoming observations is to probe the era
of cosmic reionization. Ever since the discovery that the
intergalactic gas throughout the Universe is highly ionized,
astronomers have been searching for the period when the hydrogen was
ionized for a second time after it became neutral at cosmic
recombination. This reionization is the most recent cosmic phase
transition, whereby the gas was ionized throughout the Universe,
affecting subsequent galaxy formation and potentially detectable
through its effect on a large range of observations.

\begin{figure}[htb]
\centering
\includegraphics[width=2.9in]{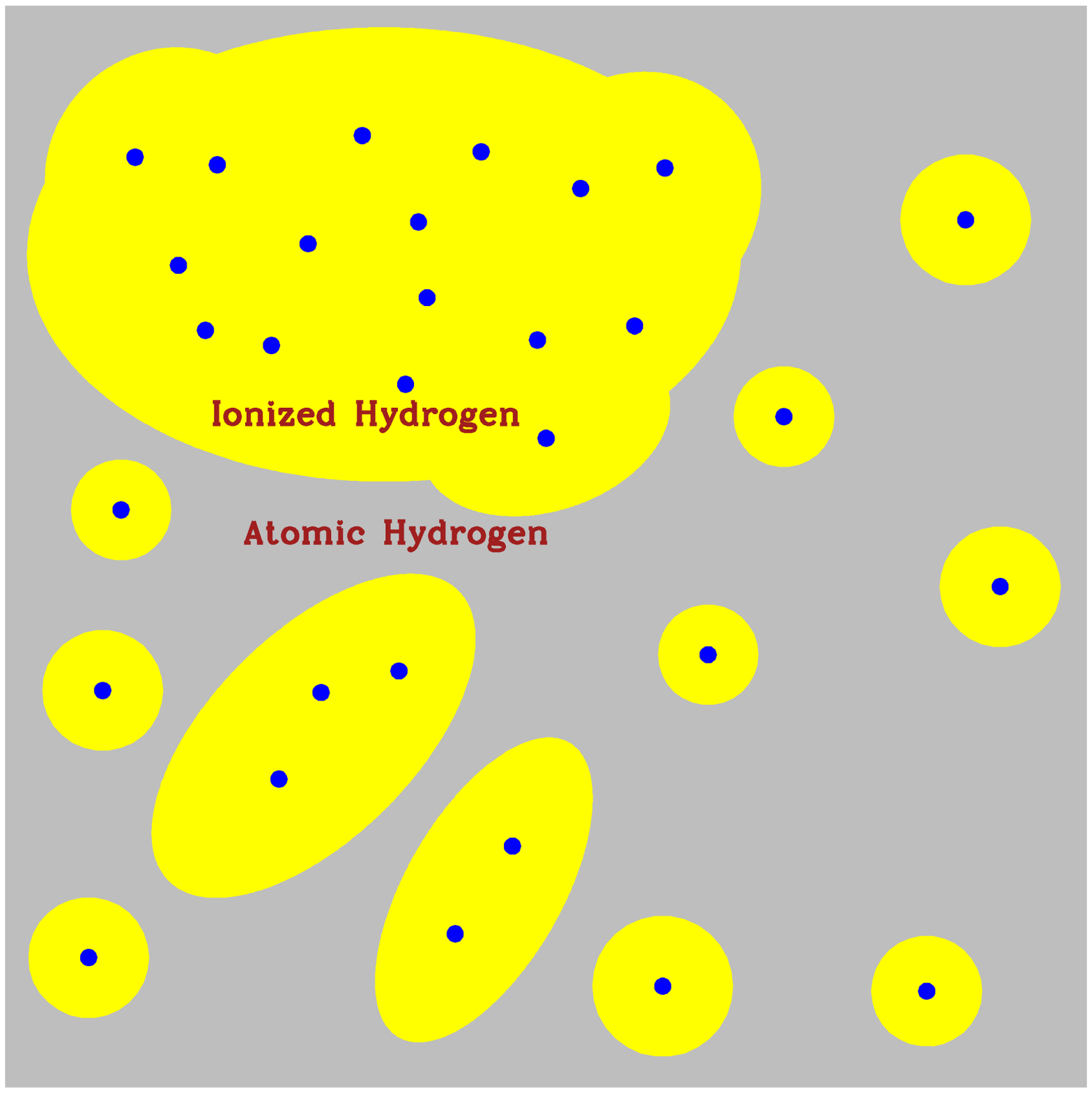}
\includegraphics[width=2.7in]{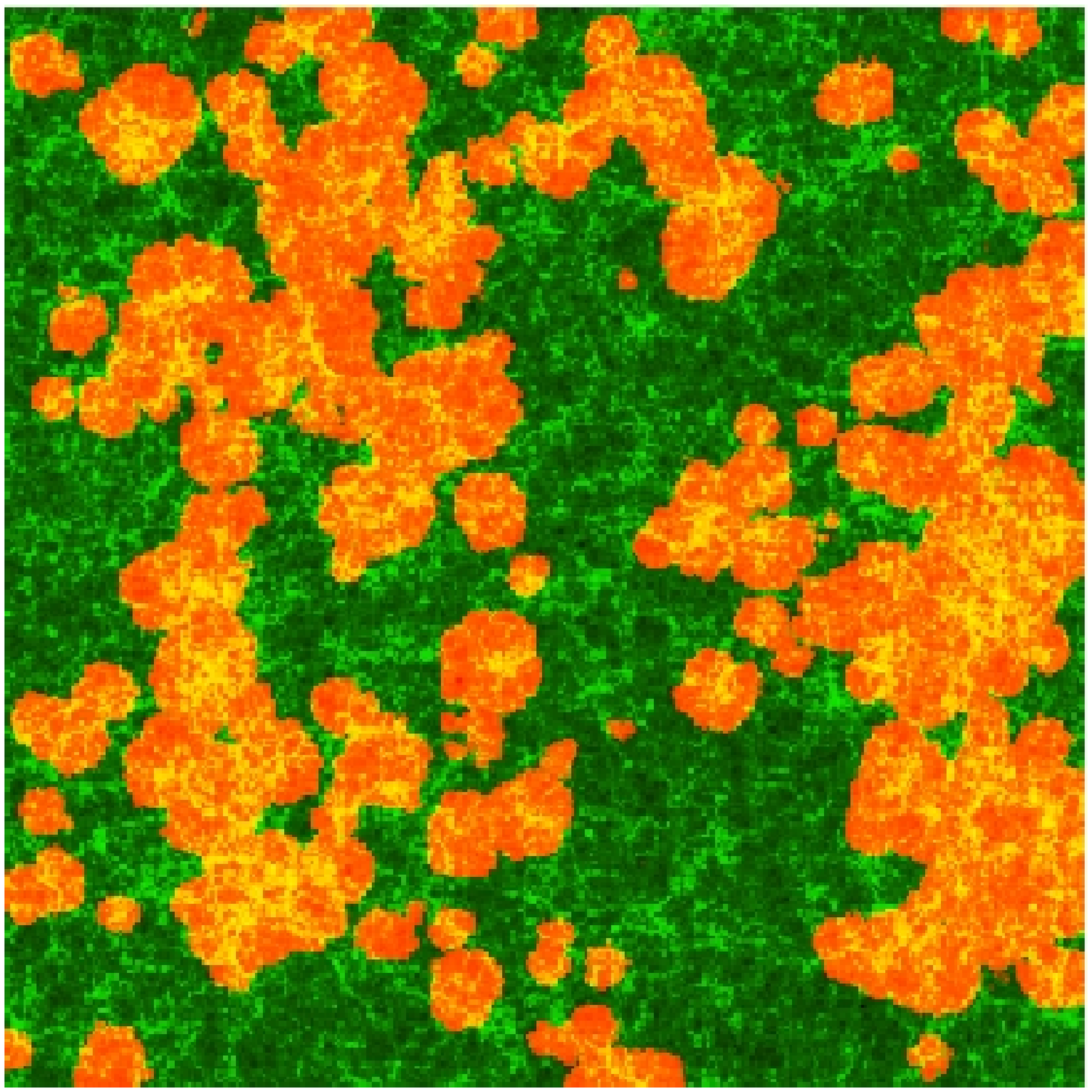}
\caption{During reionization, the ionized bubbles created by clustered
  groups of galaxies \cite{BL04} imprinted a signature in the power
  spectrum of 21-cm fluctuations \cite{fzh04}. The illustration (left
  panel, from \cite{BSc}) shows how regions with large-scale
  overdensities form large concentrations of galaxies (dots) whose
  ionizing photons produce large ionized bubbles. At the same time,
  other large regions have a low density of galaxies and are still
  mostly neutral. A similar pattern has been confirmed in large-scale
  numerical simulations of reionization (right panel, showing a
  two-dimensional slice from a 150 Mpc simulation box \cite{mellema}).
}
\label{f:reion}
\end{figure}

In 2004 we argued \cite{BL04} that the typical sizes of H~II bubbles
during reionization should be around 10 or 20 Mpc (see
Figure~\ref{f:reion}), while many numerical simulations of
reionization at the time followed a total volume below this size.
Further analytical models \cite{fzh04} and large-scale numerical
simulations \cite{Ciardi,zahn,mellema,santos} have indeed demonstrated
the dominance of large bubbles due to large groups of
strongly-clustered galaxies. This has helped motivate the large number
of observational efforts currently underway, since large-scale
fluctuations are easier to detect (as they do not require high angular
resolution).

Several arrays of low-frequency radio telescopes are currently being
constructed in order to detect the 21-cm fluctuations from cosmic
reionization. Current efforts include the Murchison Wide-field Array
(MWA), the Low Frequency Array (LOFAR), the Giant Metrewave Radio
Telescope (GMRT), and the Precision Array to Probe the Epoch of
Reionization (PAPER); plans have been made for a future Square
Kilometer Array (SKA).

In 2005 we showed \cite{BL05b} that fluctuations in the galaxy number
density cause fluctuations even in the intensity of long-range
radiation, leading to promising sources of 21-cm fluctuations at even
higher redshifts. Specifically, the 21-cm emission at high redshift is
affected by \Lya radiation from stars and X-ray radiation from stellar
remnants (mostly X-ray emitting black-hole binaries) \cite{Madau}. We
considered fluctuations in the \Lya radiation that helps maintain the
21-cm signal. Similar (though somewhat larger and later) fluctuations
are expected from heating due to an inhomogeneous X-ray background
\cite{Jonathan07}.

\section{The streaming velocity: relative motion between the baryons 
and dark matter}

Up until recently, studies of early structure formation were based on
initial conditions from linear perturbation theory. However, there is
an important effect missing from this treatment \cite{TH10}. At early
times, the electrons in the ionized gas scattered strongly with the
then-energetic CMB photons, so that the baryons moved together with
the photons in a strongly-coupled fluid. On the other hand, the motion
of the dark matter was determined by gravity, as it did not otherwise
interact with the photons. Thus, the initial inhomogeneities in the
universe led to the gas and dark matter having different velocities.
The key properties of this relative motion is that the velocity is
generated by large-scale modes, and that it contains a strong baryon
acoustic oscillation (BAO) signature.

\begin{figure}[htb]
\centering
\includegraphics[width=141mm]{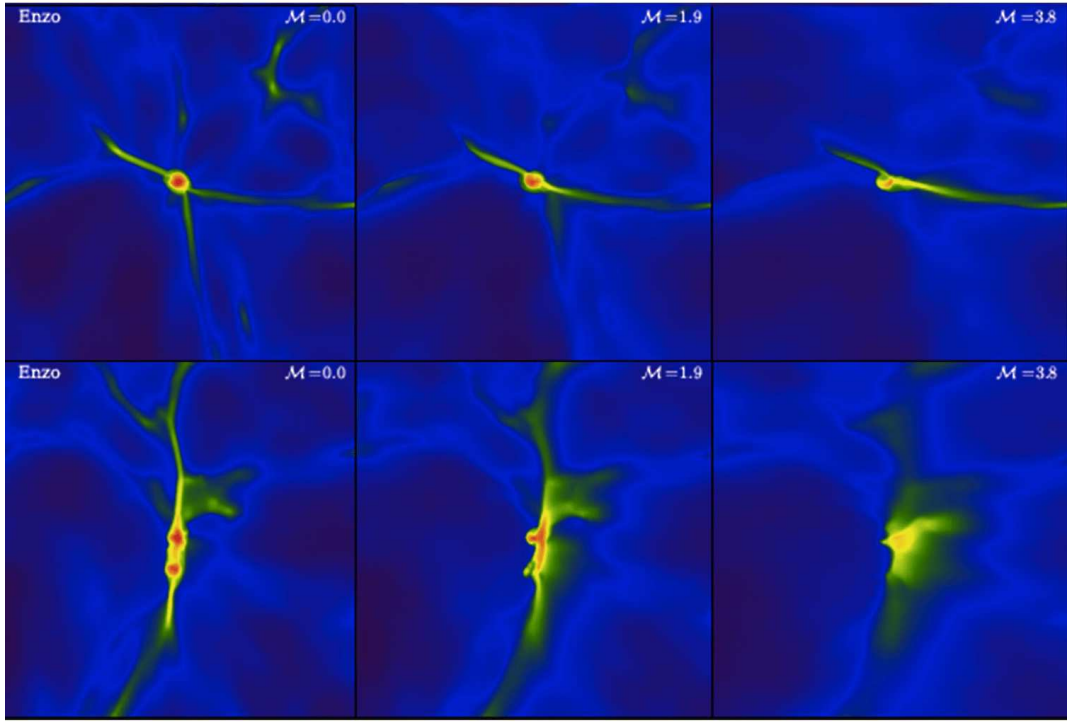}
\caption{Effect of relative velocity on individual halos, from
  numerical simulations (including gravity and hydrodynamics).  The
  colors indicate the gas density, which ranges from $10^{-26} {\rm
    g/cm}^3$ (blue) to $10^{-23} {\rm g/cm}^3$ (red). Two halos are
  shown at $z=20$, with a total halo mass of $2 \times 10^6 M_\odot$
  (top) or $8 \times 10^5 M_\odot$ (bottom). Panels show the result
  for gas initially moving to the right with a relative velocity of 0
  (left), 1 (middle), or 2 (right) in units of the root-mean-square
  value of the relative velocity at $z=20$. $\mathcal{M}$ indicates
  the corresponding Mach number at $z=20$. From \cite{mcquinn12}.}
\label{f:numsim}
\end{figure}

The streaming velocity has a major effect on the first stars, since it
suppresses the abundance of halos \cite{TH10}, suppresses the gas
content of halos \cite{Dalal}, and increases the minimum mass of halos
in which stars can form from gas that cools via molecular hydrogen
cooling \cite{Stacy:2011,Greif:2011,anastasia}. Many small-scale
numerical simulations have studied this
\cite{Maio:2011,mcquinn12,mcquinn12b,naoz,naoz2} (see
Figure~\ref{f:numsim}).

\section{Detecting the first stars at redshift 20}

Full numerical simulations that successfully resolve the mini-galaxies
that hosted the first stars are limited to $\sim 1$~Mpc volumes. On
the other hand, only large cosmological scales are accessible to 21-cm
observations (which are currently limited to low angular resolution).
Purely analytical calculations typically require simplifying
assumptions and thus achieve limited accuracy. Thus, the best way to
generate observable predictions from the era of early galaxies is with
a hybrid method that combines linear theory on large scales with the
results of numerical simulations on small scales. We recently
developed such a hybrid method and used it to produce the first
realistic, three-dimensional images of the expected large-scale
distribution of the first stars and the resulting 21-cm emission
\cite{nature}. In our approach we built upon previous hybrid methods
used for high-redshift galaxy formation \cite{TH10,Dalal,21cmfast}.

\begin{figure}[htb]
\centering
\includegraphics[width=2.9in]{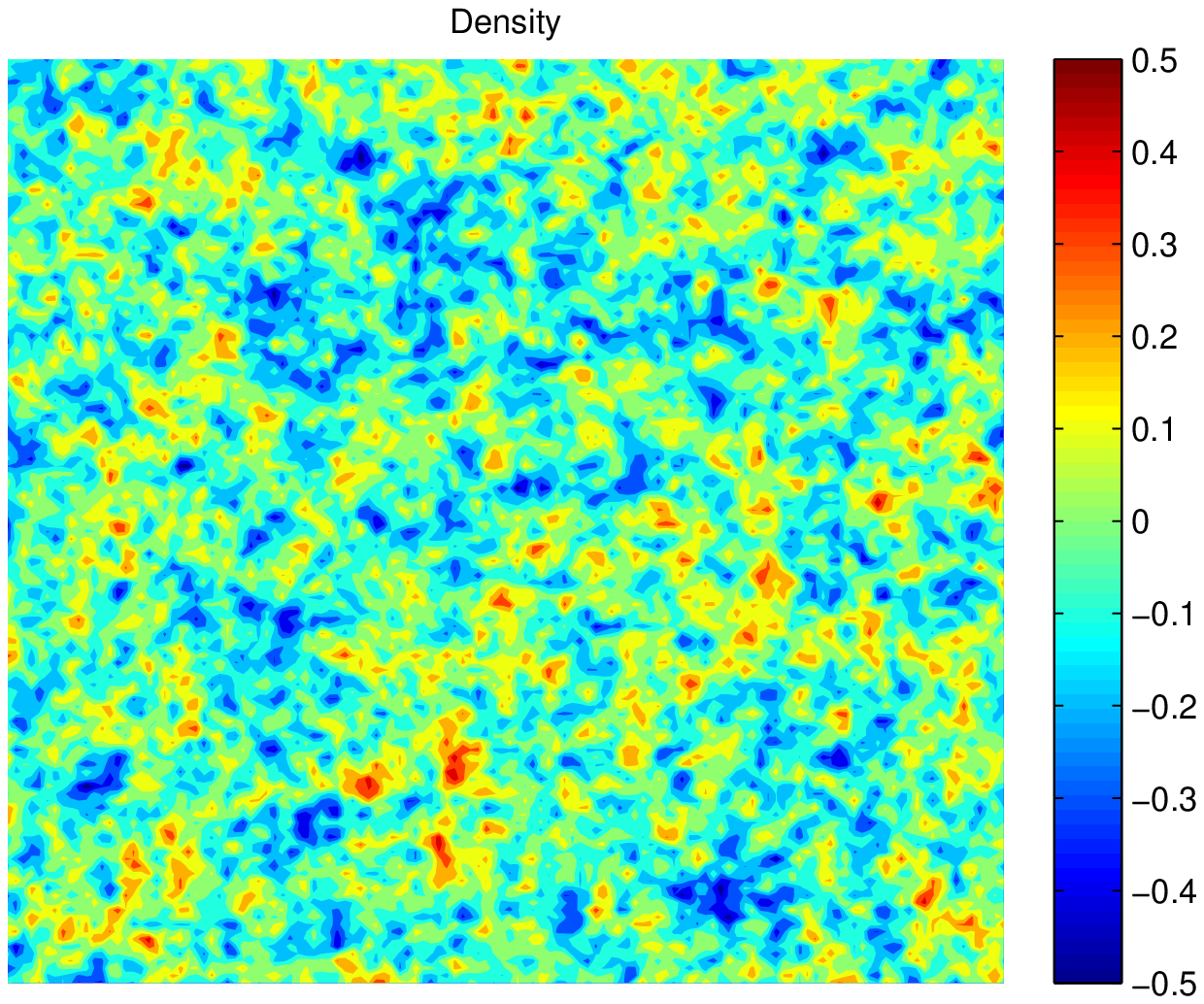}
\includegraphics[width=2.9in]{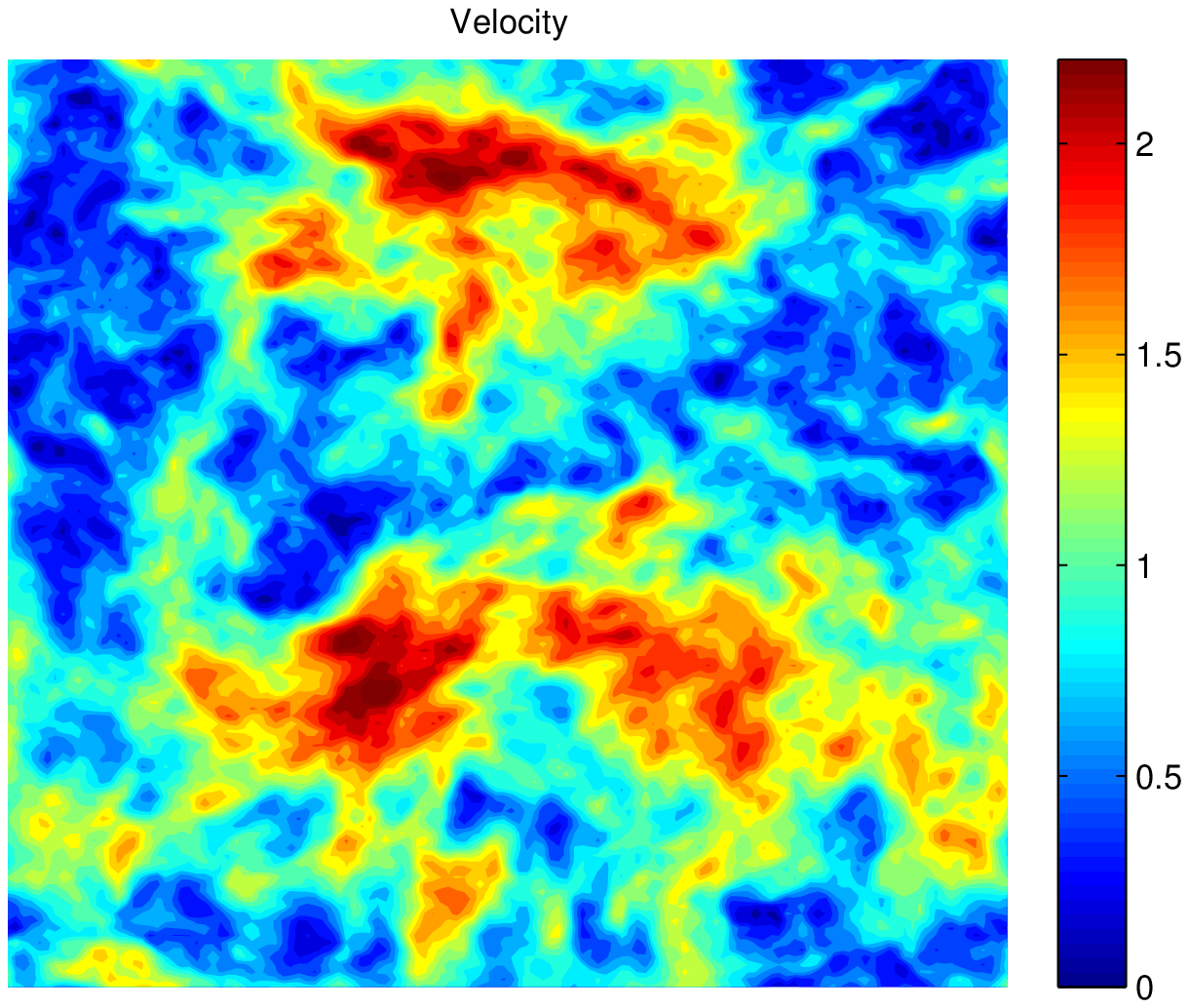}
\caption{The large-scale density and velocity fields in an example of
  a slice from a simulated volume 384 Mpc on a side (based on
  \cite{nature}, but taken from a different box from the one shown in
  the Figures in \cite{nature}, i.e., for a different set of random
  initial conditions). The thickness of the slice is 3 Mpc (which is
  also the pixel size of our grid). For the density field (left
  panel), we show the fractional perturbation relative to the mean, at
  $z=20$; for the velocity field (right), we show the magnitude of the
  relative motion in units of the root-mean-square value (the map is
  independent of redshift in these relative units).}
\label{f:RhoV}
\end{figure}

\begin{figure}[htb]
\centering
\includegraphics[width=2.9in]{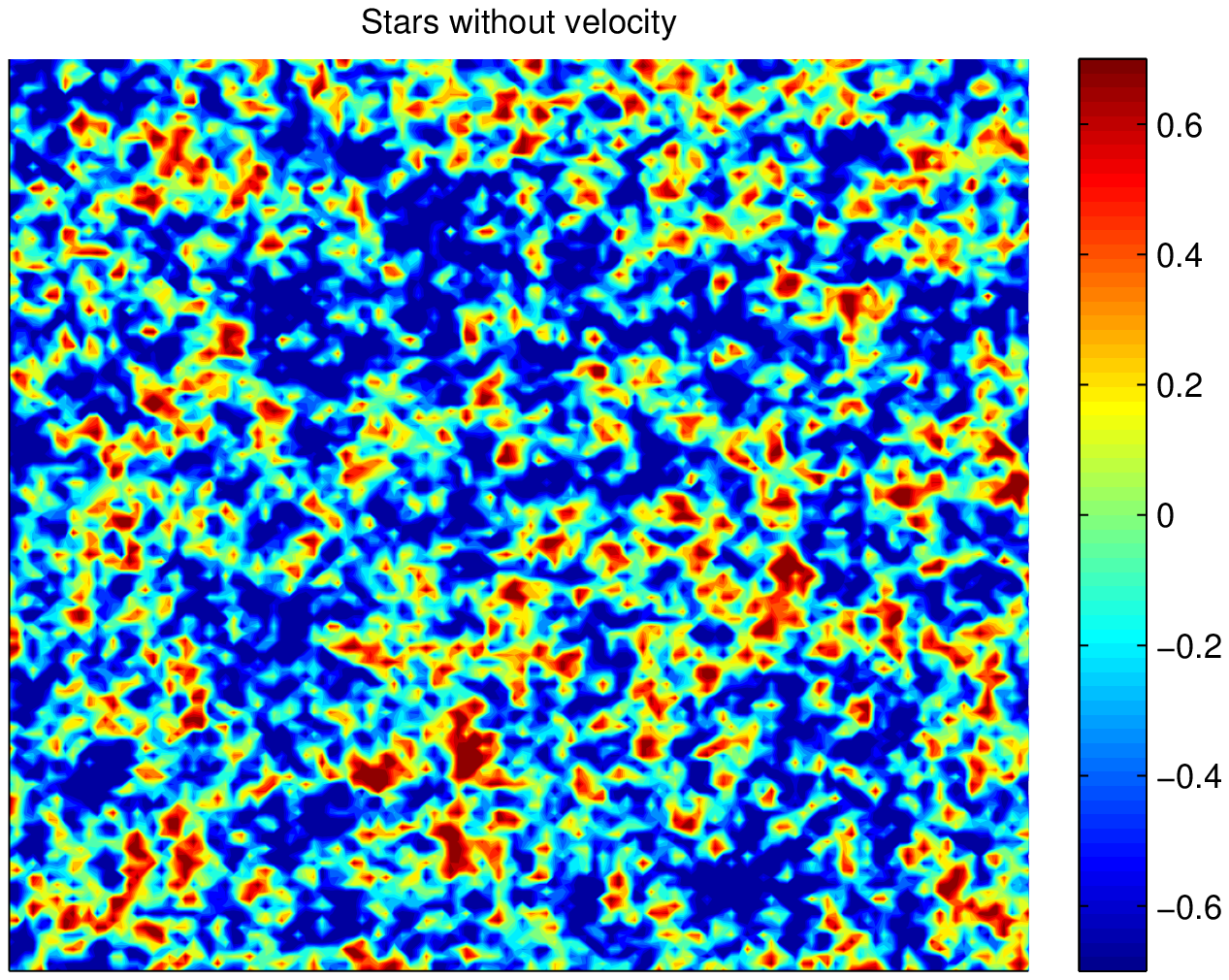}
\includegraphics[width=2.9in]{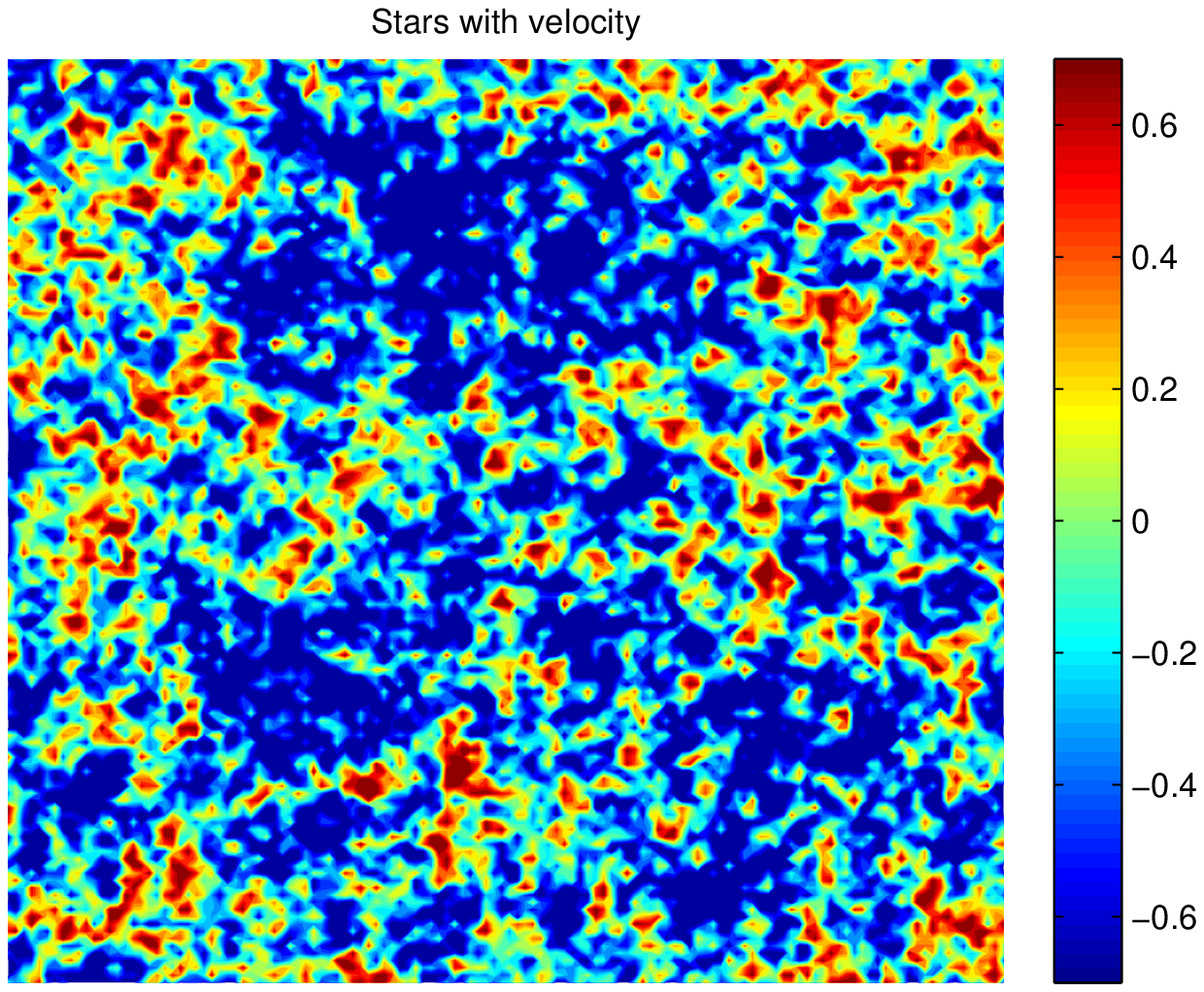}
\caption{Effect of relative velocity on the number density of stars at
  redshift 20. For the same slice as in Figure~\ref{f:RhoV}, we
  compare the previous expectation (left panel), including the effect
  of density only, to the new prediction (right), including the effect
  of the same density field plus that of the relative velocity.  The
  colors correspond to the logarithm of the gas fraction in stars,
  in units of its cosmic mean value in each case.}
\label{f:fgas20}
\end{figure}

Velocities are coherent on larger scales than the density
(Figure~\ref{f:RhoV}). The resulting distribution of stellar density
at $z=20$ is shown in Figure~\ref{f:fgas20}. The velocity effect
produces a more prominent cosmic web on large scales, marked by large
coherent regions that have a low density of stars, separated by
ribbons or filaments of high star formation. The effect is even
stronger at higher redshifts, which substantially alters the feedback
environment of the first stars.

\begin{figure}[htb]
\centering
\includegraphics[width=2.9in]{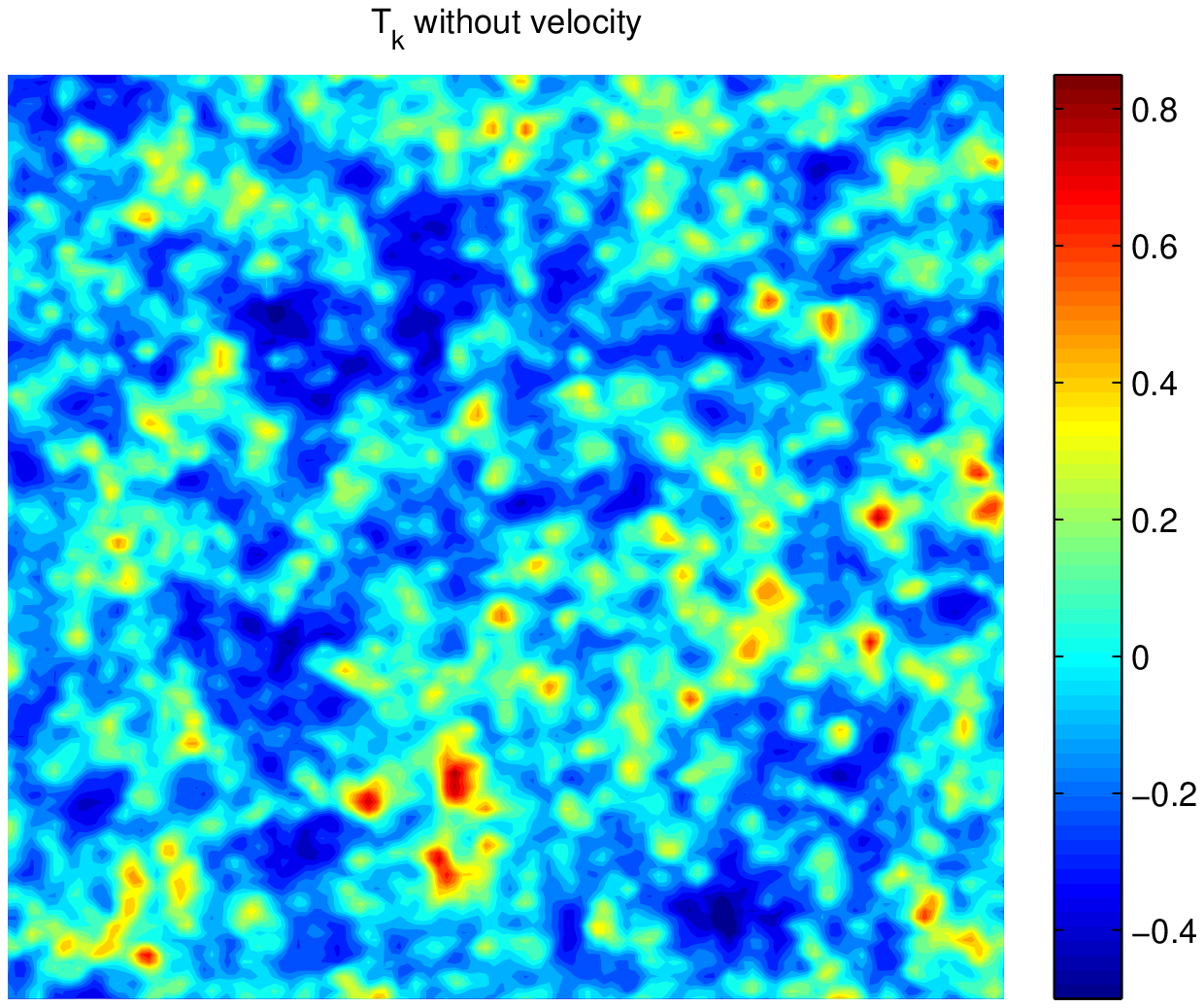}
\includegraphics[width=2.9in]{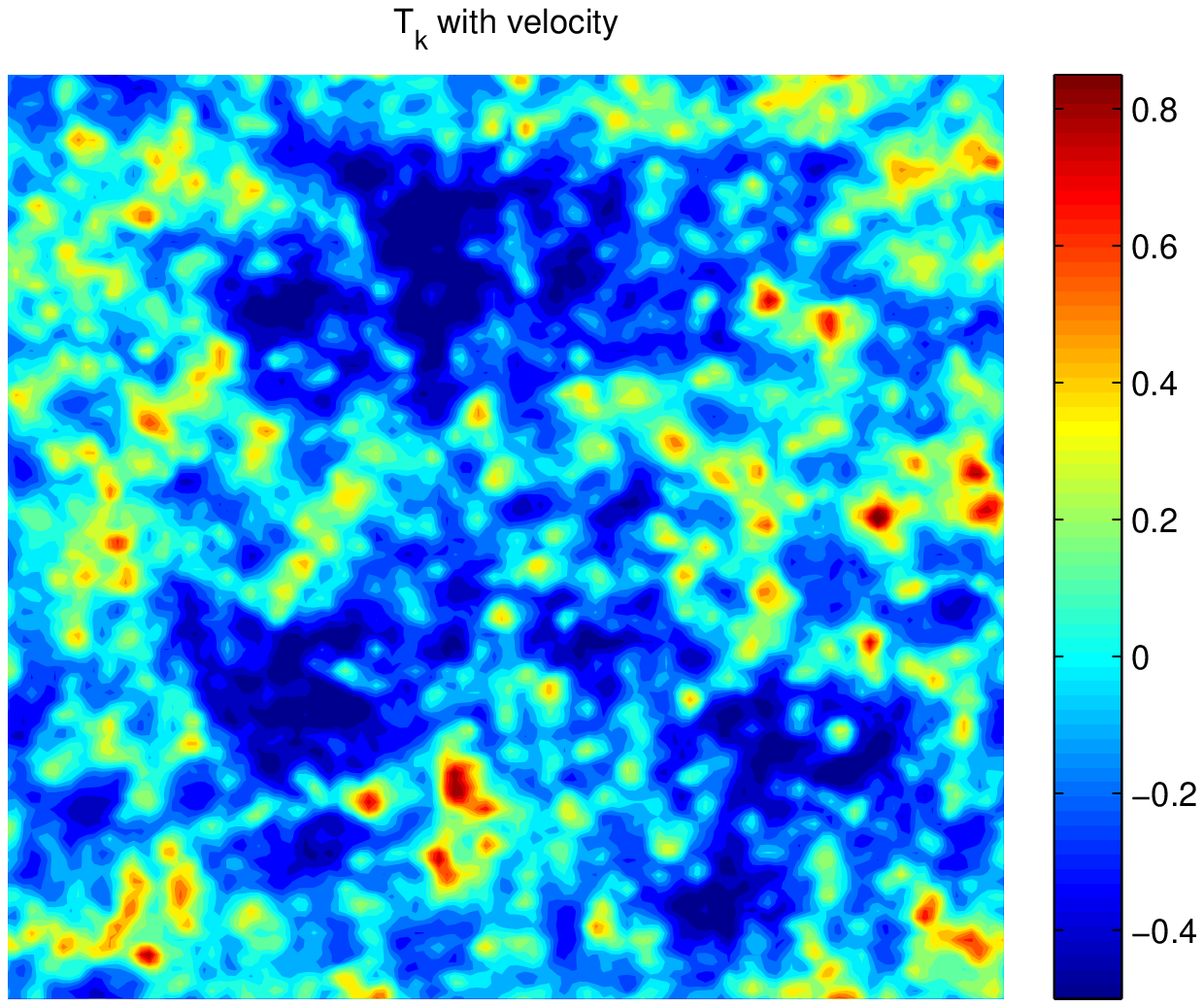}
\caption{Effect of relative velocity on the gas temperature at
  redshift 20. For the same slice as in Figure~\ref{f:RhoV}, we
  compare the previous expectation (left panel) to the new prediction
  (right), which includes the effect of relative velocity. The colors
  correspond to log of the gas (kinetic) temperature in units of the
  CMB temperature at $z=20$.}
\label{f:Tk20}
\end{figure}

\begin{figure}[htb]
\centering
\includegraphics[width=2.9in]{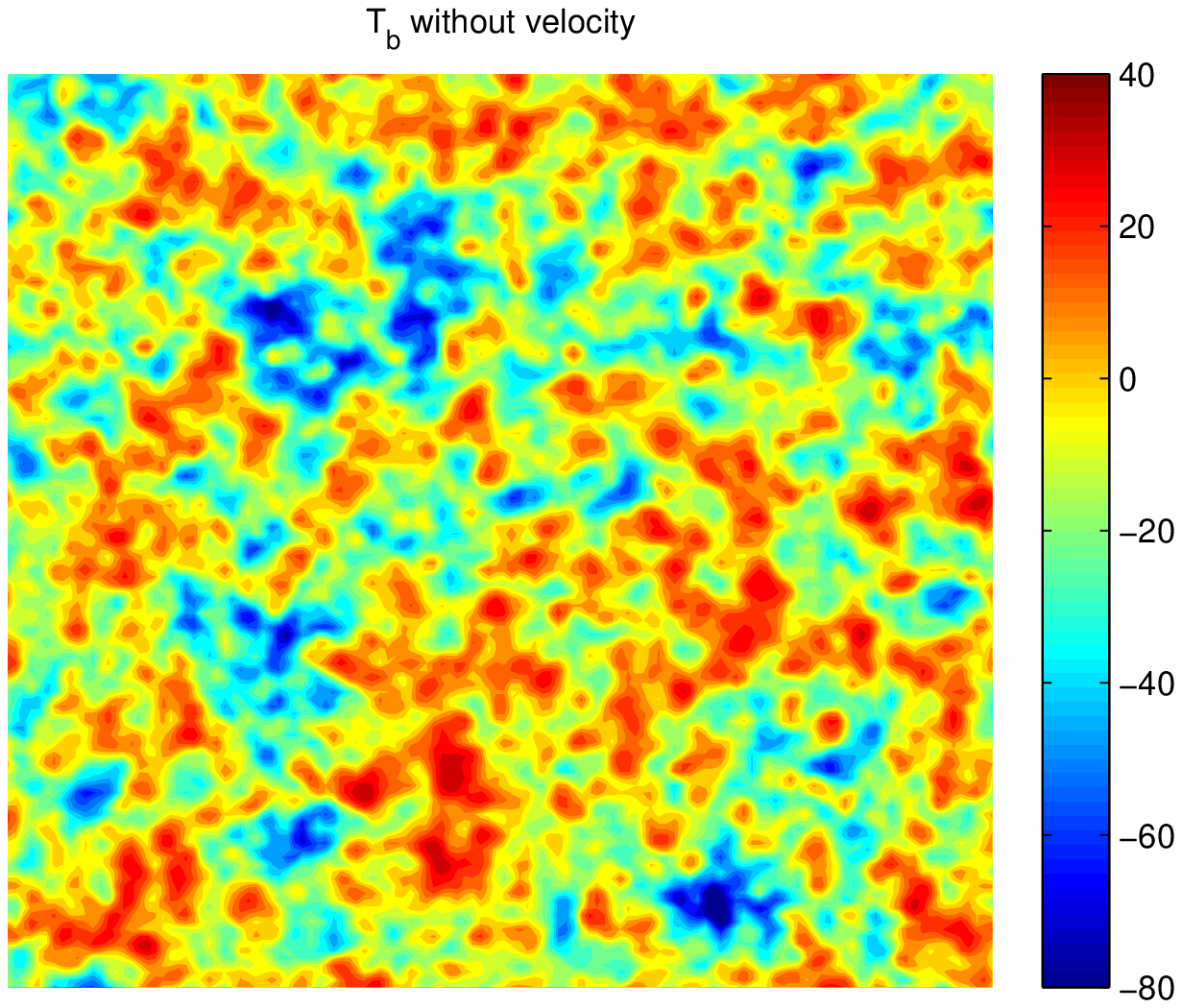}
\includegraphics[width=2.9in]{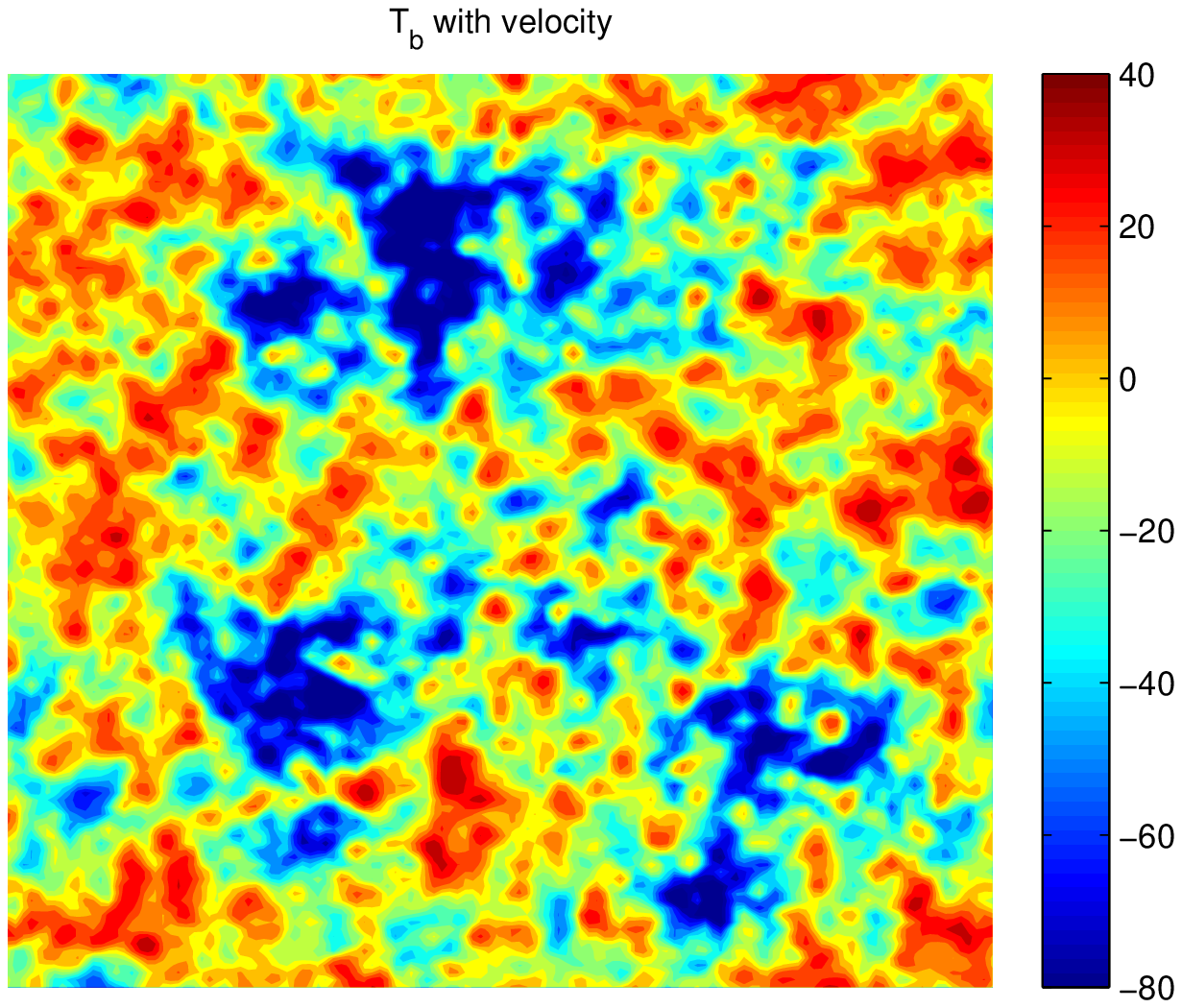}
\caption{Effect of relative velocity on the $z=20$ 21-cm brightness
  temperature (which measures the observed intensity of radio waves
  emitted by intergalactic hydrogen). For the same slice as in
  Figure~\ref{f:RhoV}, we compare the previous expectation (left
  panel) to the new prediction (right), which includes the effect of
  relative velocity. The colors correspond to the 21-cm brightness
  temperature in millikelvin units.}
\label{f:Tb20}
\end{figure}

Observationally, these degree-scale fluctuations will affect various
cosmic radiation backgrounds, and in particular the history of 21-cm
emission and absorption. We focus here on heating fluctuations, which
are maximized roughly at the heating transition (when the cosmic mean
gas temperature equals that of the cosmic microwave background (CMB)).
Figure~\ref{f:Tk20} shows the gas temperature distribution at $z=20$
(assuming the heating transition occurs at that time). Regions where
the gas moved rapidly with respect to the dark matter (dark red
regions, right panel of Figure~\ref{f:RhoV}) produced fewer stars
(dark blue regions, right panel of Figure~\ref{f:fgas20}) and thus a
lower X-ray intensity, leaving large regions with gas that is still
colder than the CMB by a factor of several (dark blue regions, right
panel of Figure~\ref{f:Tk20}). The large voids in star formation
produced by a high relative velocity lead to prominent 21-cm
absorption (dark blue regions, right panel of Figure~\ref{f:Tb20})
seen on top of the pattern from the effect of density fluctuations.
These deep 21-cm cold spots are the main observable signature of the
effect of the relative velocity on the first stars.

\begin{figure}[htb]
\centering
\includegraphics[width=141mm]{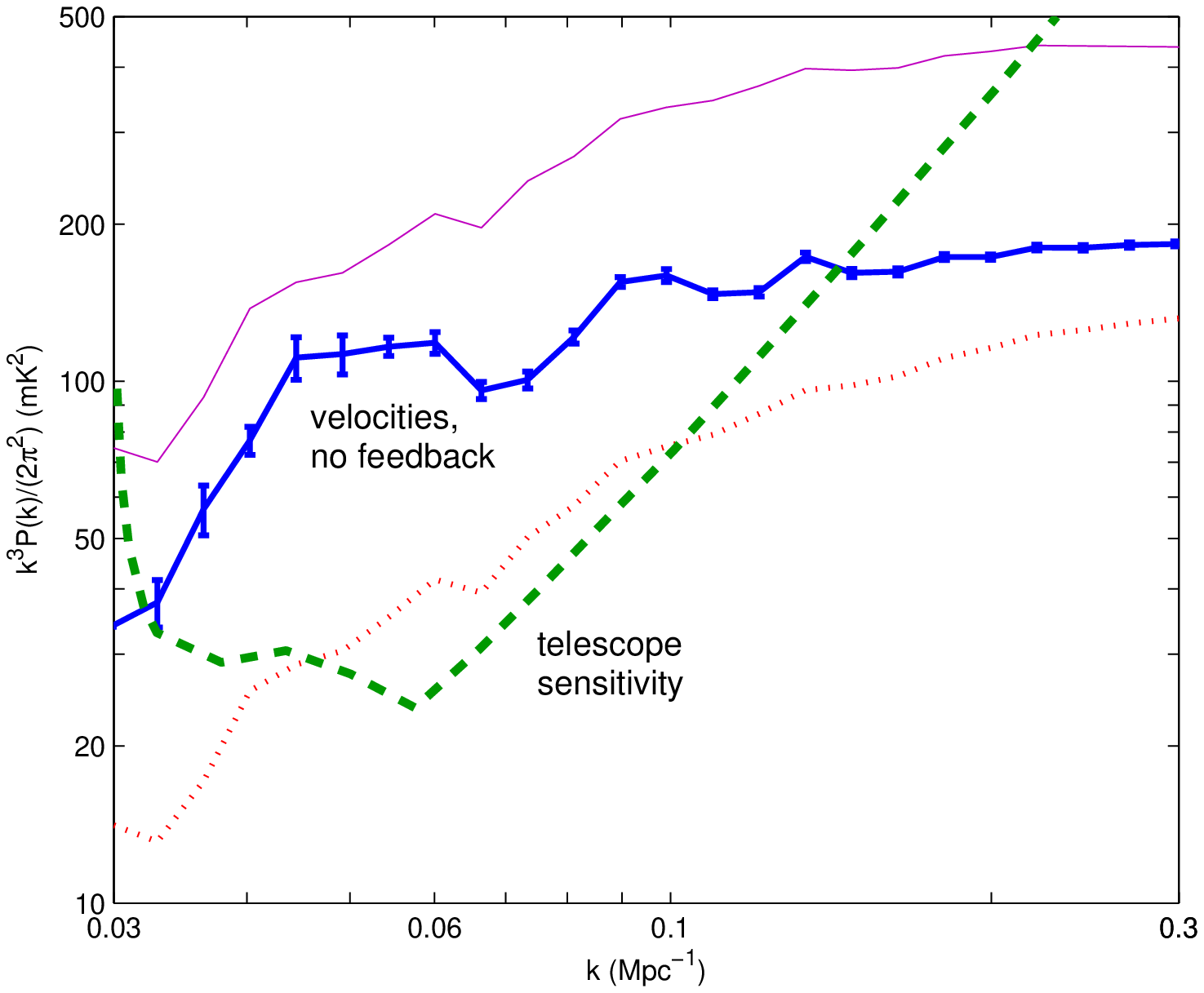}
\caption{Signature of the relative velocity in the 21-cm power
  spectrum, at the peak of the X-ray heating transition at $z=20$. We
  consider the prediction including the relative velocity effect (blue
  solid curve) or with the effect of densities only (red dotted
  curve). These predictions are compared to the projected 1-$\sigma$
  telescope sensitivity (green dashed curve) based on 1000-hour
  observations with an instrument like the MWA or LOFAR but designed
  to operate at 50--100 MHz \cite{McQuinn}. Future experiments like
  the Square Kilometer Array should reach a better sensitivity than
  this by more than an order of magnitude. To allow for the
  possibility of feedback, we also show the prediction for the case
  where star formation requires atomic cooling (purple solid curve).
  In this plot we have fixed the heating transition at $z=20$ for easy
  comparison among the various cases. From \cite{nature}.}
\label{f:nPS}
\end{figure}

\begin{figure}[htb]
\centering
\includegraphics[width=142mm]{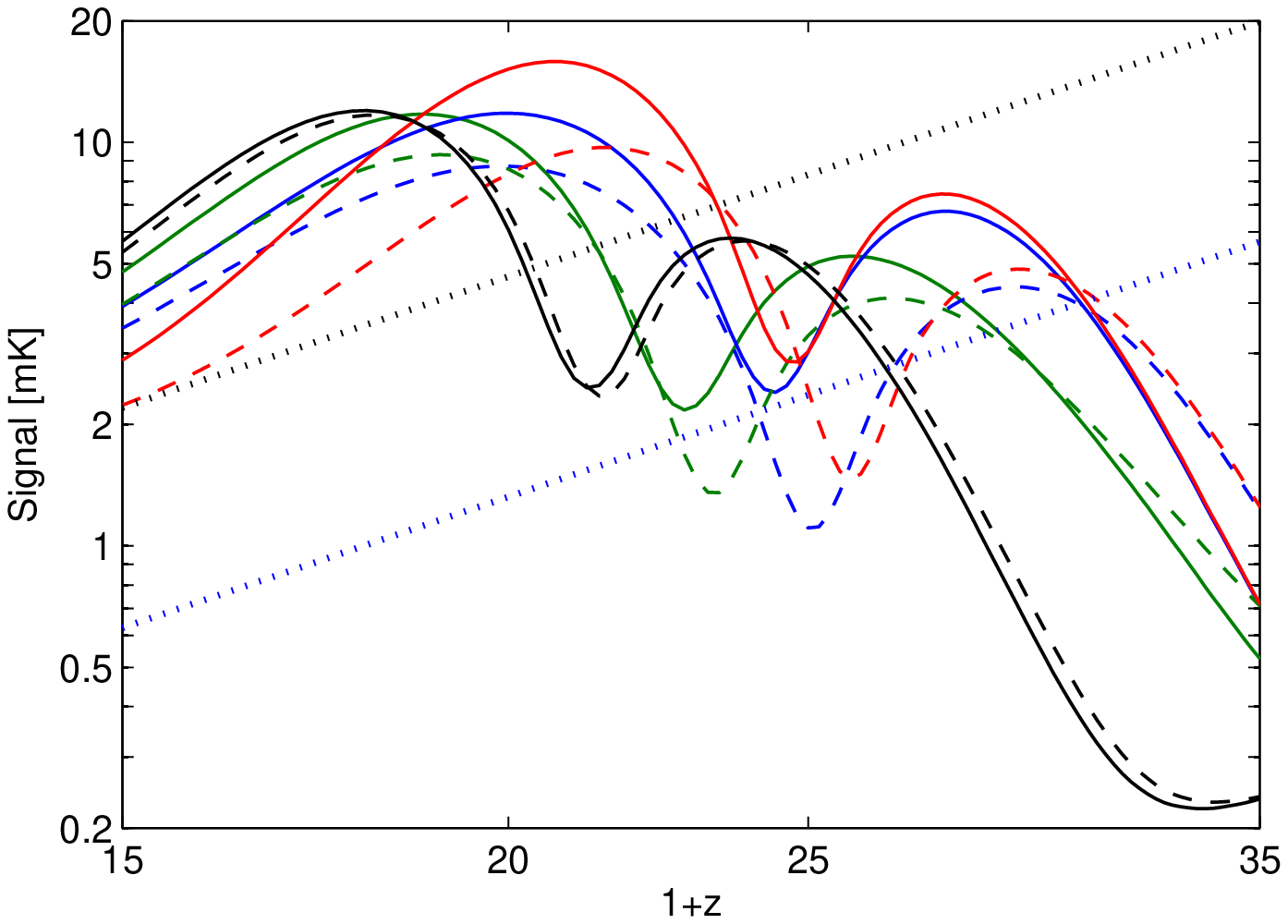}
\caption{Amplitude of the 21-cm brightness temperature at 125 Mpc, the
  BAO scale. We show the signal with (solid) and without (dashed) the
  streaming velocity for no H$_2$ cooling (black), strong (green) and
  weak (blue) feedback and for no feedback (red). We also show the
  sensitivity curves for SKA (blue dotted) and MWA/LOFAR-like (black
  dotted) experiments; the latter refers to an instrument with the
  same collecting area as MWA or LOFAR, and the former to the planned
  SKA, where to both we have applied an estimated degradation factor
  due to foreground removal (see \cite{McQuinn} and
  \cite{nature} for details). From \cite{Complete}}
\label{fig:Sig}
\end{figure}

While these Figures illustrate the detailed pattern resulting from the
effect of relative velocity on the 21-cm intensity distribution,
upcoming experiments are expected to yield very noisy maps that likely
must be analyzed statistically. Figure~\ref{f:nPS} shows the predicted
effect on a key statistic, the power spectrum of the fluctuations in
21-cm intensity (from \cite{nature}). The velocities enhance
large-scale fluctuations (blue solid curve compared with red dotted),
leading to a flatter power spectrum with prominent baryon acoustic
oscillations.  The signal is potentially observable with a redshift 20
version of current instruments (green dashed curve). At even higher
redshifts, the fluctuations imprinted by the inhomogeneous Ly$\alpha$
background in the 21-cm signal at $z \sim 25$ should be detectable
with the Square Kilometer Array (Figure~\ref{fig:Sig}).

\Acknowledgements I thank my collaborators on the cited work:
Anastasia Fialkov, Eli Visbal, Arazi Pinhas, Dmitriy Tseliakhovich,
Chris Hirata, Avi Loeb, and Smadar Naoz. (See grant acknowledgements
at bottom of cover page)

\end{document}